\documentclass[reprint,letterpaper,superscriptaddress]{revtex4-1}

\usepackage{siunitx}
\usepackage{amsmath}
\usepackage{graphicx}
\usepackage[T1]{fontenc}
\usepackage{hyperref}

\begin{document}


\title{Two-dimensional optical quasicrystal potentials for ultracold atom experiments}

\author{Theodore A. Corcovilos}
\email{corcovilost@duq.edu} 
\homepage{http://corcoviloslab.com} 
\affiliation{Duquesne University, Dept.\ of Physics, 317 Fisher Hall, 600 Forbes Ave.,  Pittsburgh, PA 15282, USA}
\affiliation{Pittsburgh Quantum Institute, B4 Thaw Hall, 4061 O'Hara St., Pittsburgh, PA 15260, USA}
\author{Jahnavee Mittal}
\affiliation{Duquesne University, Dept.\ of Physics, 317 Fisher Hall, 600 Forbes Ave.,  Pittsburgh, PA 15282, USA}


\begin{abstract}
Quasicrystals are nonperiodic structures having no translational symmetry but nonetheless possessing long-range order.
The material properties of quasicrystals, particularly their low-temperature behavior, defy easy description.
We present a compact optical setup for creating quasicrystal optical potentials with 5-fold symmetry   using interference of nearly co-propagating beams for use in ultracold atom quantum simulation experiments.
We verify the optical design through numerical simulations and demonstrate a prototype system.
We also discuss generating phason excitations and quantized transport in the quasicrystal through phase modulation of the beams.
\end{abstract}
\maketitle
\section{Introduction}
Quasicrystals (QCs) are solids, typically ternary metallic alloys, that lack translational symmetry but still exhibit long range order through a deterministic non-periodic tiling structure, leading to unusual geometric properties such as 5-, 8-, 10-, or 12-fold rotational symmetry and scale invariance\cite{Janot2012,Janssen2018}.
Such non-periodic tilings have been studied by mathematicians, most notably in the 1970's by Penrose\cite{Gardner1977}, and have existed for centuries in art and architecture\cite{Lu2007},
but material examples were unknown until the serendipitous synthesis and discovery of QCs in an AlMn alloy by Shechtman was announced in 1984\cite{Shechtman1984}, for which he was awarded the 2011 Nobel Prize in Chemistry.
QCs have also been found naturally occurring in meteorites\cite{Bindi2009},
and form spontaneously in some organic polymer systems\cite{Hayashida2007}.

Our motivation is to study 
two-dimensional quasicrystals to better understand their physical properties, including topological structure, and inform potential applications 
through quantum simulation of these materials using ultracold atoms\cite{Bloch2005,Bloch2007,Sanchez-Palencia2018}.
This Research Article presents our method of generating the quasicrystal optical potentials into which we will load quantum degenerate or nearly degenerate dilute atomic gases.

Current atomic physics studies of two-dimensional crystalline systems have been mostly limited to systems that can be generated by the interference of a small number of laser beams, such as square or hexagonal lattices\cite{Jo2012,Tarruell2012}, generated by the projection of digitally generated images\cite{Gauthier2016} and holographically projected potentials\cite{Pasienski2008,Gaunt2012,Zupancic2016},
or created by optical tweezer arrays\cite{Nogrette2014,Barredo2016,Endres2016,Barredo2018}.
We instead use the interference of nearly co-propagating lasers\cite{Nelson2007} to generate non-periodic quasicrystal potentials with 5-fold rotational symmetry.
Key advantages of our method over those above are its compactness and simplicity.

This Research Article is structured as follows.
We begin with a discussion of the known physical properties of quasicrystals and summarize the advantages of using ultracold atoms to model such systems (Sec.~\ref{sec:background}).
We then explain how interference can be used to generate quasicrystal optical potentials and present numerical simulations of our proposed optical system (Sec.~\ref{sec:numerical}).
In Sec.~\ref{sec:mask}, we build a simplified optical system based on a shadow mask and show that it indeed generates the required potential.
We conclude in Sec.~\ref{sec:mod} by introducing a novel type of quasicrystal excitations, phasons, that are inaccessible in material samples but that can be generated with our setup.
These excitations can elucidate the topological structure of these materials.

\section{Background and motivation}\label{sec:background}
Quasicrystals are often described as existing between the extremes of ordered crystals and amorphous materials\cite{Janot2012,Janssen2018}.
Even when their constituent atoms are metallic, pure QCs show bulk properties distinct from those of crystalline metals and metallic glasses.
Empirically, QCs act as a chimera of metals, nonmetals, and semiconductors.
The key phenomenological observations of QCs were established in the 1990's\cite{Biggs1990,Boissieu1993,Janot1993,Pierce1993,Mayou1993,Lindqvist1993}, but progress in understanding the underlying quantum states has slowed.
Theoretical approaches to understanding QCs are difficult because of the nonperiodic geometry.

The primary unresolved issue that we will seek to address is the nature of the low-temperature localized state.
Two-dimensional crystalline systems show a superfluid to Mott insulator transition at low temperatures and integer filling factors.  This has been verified using time-of-flight imaging, noise-correlation measurements\cite{Spielman2007}, and in-situ imaging to determine compressibility\cite{Gemelke2009}.
Systems with disorder, however, exhibit Anderson localization\cite{Anderson1958,Abrahams1979} caused by the destructive interference of single-particle wavefunctions over multiple random paths.
The experimental signature of Anderson localization is the exponential decay of particle momentum distributions\cite{Billy2008,Chen2009,Choi2016}.
Some quasi-periodic systems exhibit weaker localization than disordered systems\cite{Lueschen2016,Bordia2017,Pilati2017}, and quasicrystals are also expected to show this behavior.
Furthermore, interactions between particles, such as the contact interactions between bosonic atoms that we use, are predicted to induce a variety of many-body states\cite{Hou2018}. 

One challenge of studying quasicrystal systems is the difficulty in generating viable numerical predictions to compare with.
Typical techniques for solving Schr\"odinger's equation on crystalline systems (e.g.\@ Bloch's theorem, Wannier basis functions, or periodic boundary conditions) do not apply to our system because of the nonperiodic structure of the quasicrystals\cite{Janot2012,Janssen2018}.
Direct approaches such as diagonalization of the full Hamiltonian are computationally intractable because of the large basis required to fully capture the structure.
Although some one-dimensional nonperiodic systems have been solved\cite{Kohmoto1987,Schreiber2015,Singh2015}, few two-dimensional systems have been solved\cite{Grimm2003,Jagannathan2007,Fuchs2016,Bandres2016}.
As such, solving for the single-particle and many-particle eigenstates of the quasicrystal potential is beyond the scope of this article.
Indeed, inspiring theorists and computationalists to take on this challenge is a motivating factor of our experiments.

Because calculations of quasicrystals are difficult,
we approach the system by studying an analogous quantum mechanical system with similar geometry.
This geometry will be implemented as an optical potential generated by interference.
In a typical optical lattice, several laser beams of the same wavelength interfere with each other to create alternating stationary regions of high and low intensity.
Optical lattices have been the cornerstone of quantum emulation experiments for the past two decades\cite{Bloch2005,Bloch2007,Lewenstein2007,Daley2018,Sanchez-Palencia2018}.
The optical lattice plays the role of the ionic potential landscape in which the atoms, playing the role of electrons, live.
We expand this technique to nonperiodic lattices.
Earlier work on simulating quasicrystals with cold atoms used planar arrangements of interfering beams\cite{Guidoni1997,Guidoni1999,Sanchez-Palencia2005,Cetoli2011,Viebahn2018} or looked at one-dimensional systems\cite{Singh2015,Mace2016}. 
Our technical approach, described below, uses a compact setup requiring minimal optical access to the sample and gives us flexibility in choosing the size of the lattice spacing in the quasicrystal, which we use to match the lattice spacing to the resolution of our imaging system.
We will also discuss 
various excitation modes in the quasicrystal, some of which are unique to these structures.

\section{Generating quasicrystal optical potentials}\label{sec:numerical}

Our procedure for generating the quasicrystal optical potentials is described below, 
along with simulations to show its feasibility.
A low-power prototype using this scheme is presented later in Sec.~\ref{sec:mask}.
The key element is a particular arrangement of optical fibers that will emit laser beams that interfere to generate the nonperiodic lattice.
The technical requirements of the optical potential are
(1) maximal overlapping of the constituent beams,
(2) accurate setting of the polarization to ensure maximum contrast in the interference pattern, and
(3) phase stability between the beams to prevent unwanted ``phason'' excitations in the lattice.
These points will be addressed below.
With these requirements met, we generate an optical potential with a characteristic scale of $\sim \SI{2}{\micro\meter}$ at the plane of the atoms. 

\subsection{Interference of multiple plane waves}

To motivate our approach, we first consider the behavior of interfering plane waves.
The electric field $\mathbf{E}$ of a set of $N$ monochromatic plane waves is given by their superposition:
\begin{equation}
\mathbf{E}(\mathbf{r},t) = \sum\limits_{j=1}^{N}
E_j \vec{\epsilon_j} e^{i(\mathbf{k}_j\cdot\mathbf{r}-\omega t + \phi_j)},
\end{equation}
where $\mathbf{r}$ is position, $t$ is time, $E_j$ is the electric field amplitude of the $j$-th beam, $\vec{\epsilon}_j$ is the polarization unit-vector, $\mathbf{k}_j$ is the propagation wavevector with magnitude $k$, $\omega=ck$ is the angular frequency of the light, and $\phi_j$ is the relative phase of the beam.
We assume the light is propagating through vacuum with speed $c$.
The physical electric field is equal to the real part of the above expression.
The irradiance corresponding to this electric field is
\[
I(\mathbf{r}) = \frac{1}{2Z_0} \left< \left|\mathbf{E}{(\mathbf{r},t)} \right|^2 \right>,
\]
where $Z_0\approx \SI{377}{\ohm}$ is the impedance of the vacuum, and the angle brackets indicate time averaging.

If all of the beams have equal amplitude $E$ and polarization, then the irradiance becomes
\begin{align}
\notag
I(\mathbf{r}) &= \frac{E^2}{2Z_0} \sum\limits_{j,\ell}
e^{i\left((\mathbf{k}_j-\mathbf{k}_\ell)\cdot\mathbf{r}+(\phi_j-\phi_\ell)\right)}\\
\label{eq:Isum}&=\frac{E^2}{2Z_0} \left(N +2 \sum\limits_{j<\ell} \cos\Big((\mathbf{k}_j-\mathbf{k}_\ell)\cdot\mathbf{r}+(\phi_j-\phi_\ell)\Big)\right)
\end{align}
Taking the Fourier transform of $I(\mathbf{r})$ shows peaks at zero spatial frequency and at the differences of the input $\mathbf{k}_j$ vectors.
For $N$ beams there will be $N(N-1)$ Fourier peaks in addition to the DC peak.

As a simple example, consider a pair of intersecting beams separated by an angle $2\alpha$, traveling in the $xy$ plane.
We will use such as system, which we call a ``pancake'' lattice, in our experiment to generate a 1D lattice, one period of which contains the plane of our quasicrystal samples (details below in Sec.~\ref{sec:beam}).
This provides the confinement normal to the sample plane.

The two beams travel with wavevectors
$\mathbf{k}_{\pm} = \frac{2\pi}{\lambda} (\cos\alpha\,\hat{x}\pm \sin\alpha\,\hat{y}) $
and are polarized in the $\hat{z}$ direction, where $\lambda$ is the wavelength of the light.
Assuming the plane waves have equal amplitude $E$ and a phase difference of $\phi$ between them,
the resulting electric field and irradiance are

\begin{align}
\notag\mathbf{E}(x,y,z,t) &= 2E\,\hat{z}\;e^{i(k \cos\alpha\;x-\omega t)}
\cos(k \sin\alpha\,y + \phi/2) \\
\label{eq:pancakeI}
I(x,y,z) &
= \frac{E^2}{Z_0} \bigl(1+ \cos(2k\sin\alpha\;y + \phi) \bigr)
\end{align}

The cosine term indicates a standing wave in irradiance with wavelength $\lambda/(2\sin\alpha)$ and wavevector along $\hat{y}$.  
Changing the phase $\phi$ translates the standing wave along the $y$ direction.
Note that in the limits of co-propagating ($2\alpha=0$) and counter-propagating ($2\alpha=\pi$) beams, we get the familiar results of simple superposition and a standing wave of wavelength $\lambda/2$, respectively.

Expanding the system to more beams yields increasingly more complex standing wave patterns. 
Three and four beams yield periodic 2-dimensional lattices in the $yz$ plane\cite{Jo2012,Tarruell2012}, but five interfering beams generate quasicrystals.  
Previous experiments have generated nonperiodic potentials using the interference of multiple intersecting lasers with all of the $\mathbf{k}_j$ in the same plane\cite{Guidoni1997,Sanchez-Palencia2005,Mace2016,Viebahn2018}.
Instead, we will use five nearly co-propagating beams to generate our quasicrystal potential, similarly to how we use two nearly co-propagating beams in the example above.
Our reasons for using this geometry are (1) the ability to choose the lattice scale by adjusting the angle $\alpha$, specifically to match the resolution of our imaging system, and (2) the compactness of the optical system -- there is in effect only one (composite) beam path rather than separate paths for each beam.

\subsection{Optics design}\label{sec:beam}
\begin{figure}[ht]
	\centering%
	\includegraphics[height=1.8in]{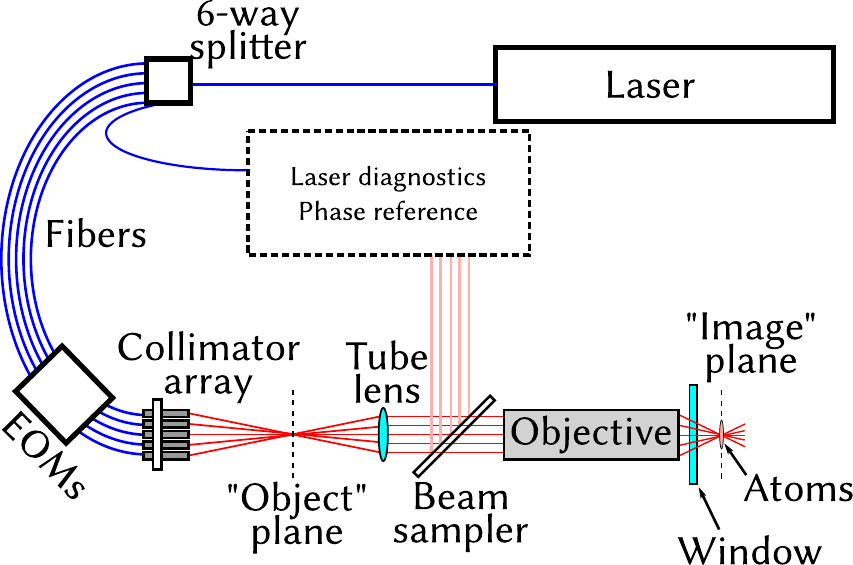}\\[1.5ex]%
	\includegraphics[height=1.8in]{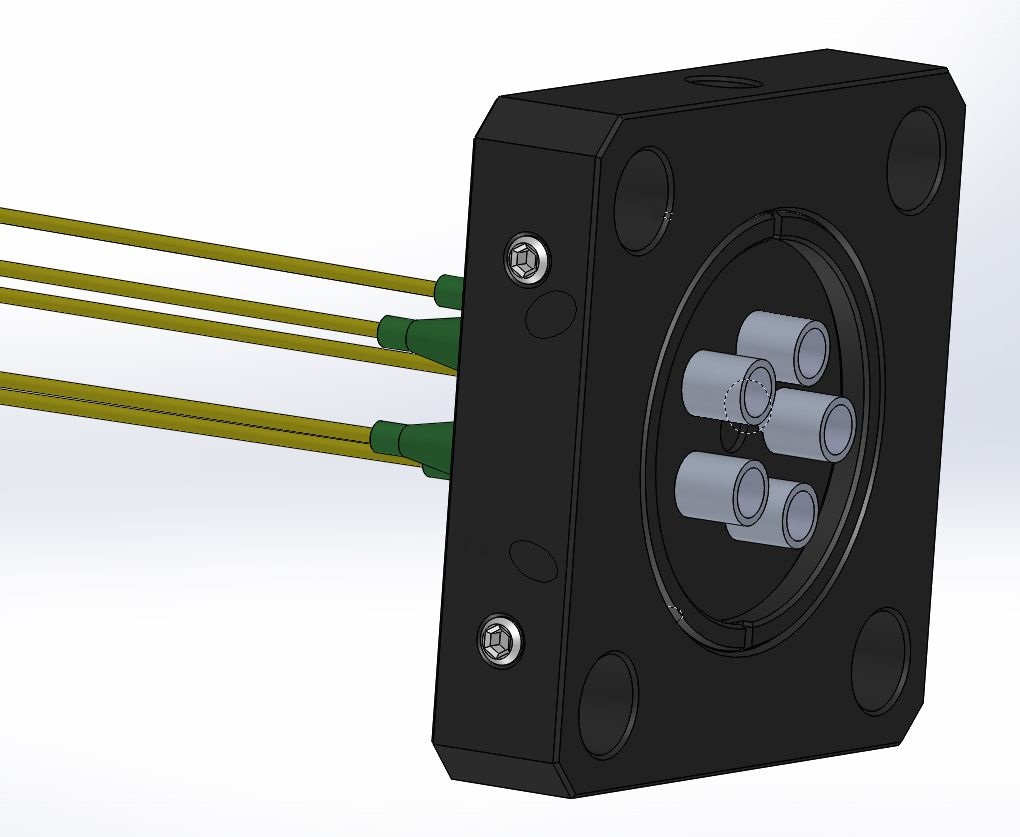}%
	\caption{Drawings of the optical system for generating the quasicrystal potentials.
		(top) System layout, not to scale.
		Light from the laser is transmitted through an optical fiber (blue) to an integrated 1-to-6 beamsplitter.
		Five of the output fibers end in collimation lenses, and the sixth goes to diagnostics.
		The intersecting beams (red) from the collimator array form an interference pattern in the object plane which is demagnified $50\times$ by the reversed microscope before being projected onto the image plane where the atoms reside.
		Electro-optic modulators (EOMs) inline with the fibers allow stabilization of the phases of the individual beams or modulation of the phases.
		(bottom) CAD detail of the optical fiber collimators and mount.
		The output beams are tilted towards the common axis by 10 milliradians.
		For scale, the diameter of the mounting plate is 25.4 mm.%
		\label{fig:layout}}
\end{figure}

The optical layout is summarized in Fig.~\ref{fig:layout}.
The light source is a set of five $\mathrm{TEM}_{00}$ laser beams emitted from optical fiber pigtail collimators. 
The individual beams have a wavelength of $\lambda=\SI{850}{nm}$, beam radius ($1/e^2$) of $\SI{0.2}{mm}$, beam divergence half-angle of $\SI{1.4}{mrad}$, and are circularly polarized.
The collimators (modeled from OZ Optics LPC-01-2AS\cite{OZ}) are arranged equidistantly on a circle of radius $\SI{5}{mm}$ and tilted towards the common axis by $\alpha=\SI{10}{mrad}$ (Fig.~\ref{fig:layout}, lower panel). 
Similar devices with fiber bundles have been used in the past to generate optical patterns\cite{Tam2004}, however those devices image the output fiber array rather than utilizing interference.

The beams intersect at a distance of 510 mm.
At this location, the quasicrystal pattern appears with a lattice scale on the order of $\lambda/\alpha \sim \SI{100}{\micro\meter}$.
We demagnify this pattern by $50\times$ by sending the beams backwards through the same microscope used to image the atoms, projecting the interference pattern onto the plane of the atoms.
The lattice scale is chosen to match the imaging resolution of the microscope objective (Mitutoyo G Plan Apo 50 \cite{Mitutoyo2012}) of $\sim\SI{2}{\micro\meter}$ (Rayleigh criterion).

The confinement normal to the sample plane ($z$ axis) will be provided by an additional 1D lattice.
Our system will use two crossed 2-W beams at $\lambda_z=\SI{1064}{nm}$ with a $1/e^2$ radius
of \SI{50}{\micro\meter} and crossing half-angle of $\alpha_z=5^\circ$, resulting in an optical lattice spacing of $\lambda_{\textrm{lat.}}=\lambda_z/(2\sin\alpha_z) = \SI{6}{\micro\meter}$ (importantly, larger than the depth-of-focus of the imaging objective of \SI{2}{\micro\meter}, so that only one plane is in focus during data acquisition) and a trap depth of $U_z/k_B \approx \SI{-1700}{\micro\kelvin}$,
where $k_B$ is Boltzmann’s constant (Eq.\eqref{eq:pancakeI}).
Approximating the bottom of the potential as a harmonic oscillator yields an energy spacing of $\hbar\omega_{\textrm{h.o.}}/k_B \approx \SI{2.3}{\micro\kelvin}$.
This is about 30 times the expected temperature of the atoms, so the $z$ degrees of freedom are essentially frozen out, yielding an effectively two-dimensional system.

\subsection{Simulation results}
To show that our procedure for generating and modulating quasicrystal potentials is effective, we present a series of numerical simulations using realistic laboratory parameters.  Later, in Sec.~\ref{sec:mask} we demonstrate a low-power prototype of the optical system.
We numerically simulate the system described above 
using the \textsc{poppy} wave optics software package\cite{Perrin2012,Perrin2017} for Python.  
The simulation uses Fresnel propagation of scalar waveforms in the paraxial approximation and includes all of the lenses (approximated as diffraction-limited thin lenses) and apertures in the path from the fibers to the atoms.
Diffraction effects are included in the calculation for completeness 
resulting in weak Fresnel rings of about 10\% modulation depth in irradiance.

\begin{figure}
	\centering%
\includegraphics{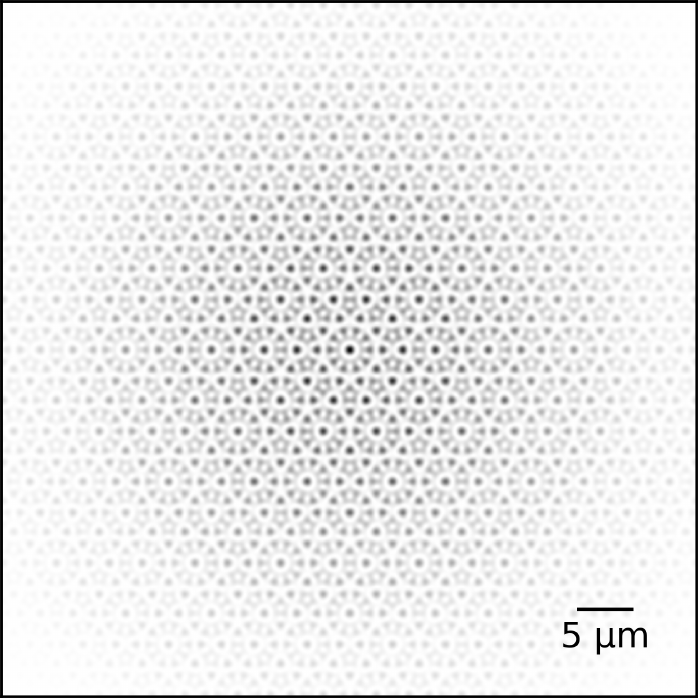}\\
\includegraphics{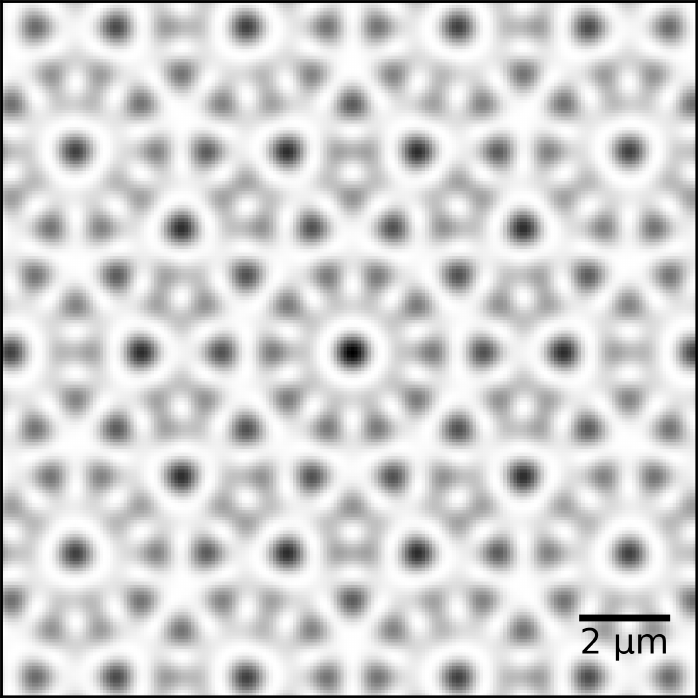}
	\caption{Optical potential numerical simulation results.
	Darker pixels represent the deepest regions of the potential.
		(top) Optical dipole potential seen by the atoms, darker regions indicate 	potential wells.
		(bottom) Detail of the center of the potential, showing the length scale and the 5-fold quasicrystal rotational symmetry.
		For the irradiance scale, see Fig.~\ref{fig:cuts}.
	}\label{fig:QCsim}
\end{figure}

\begin{figure}
	\centering%
	\includegraphics{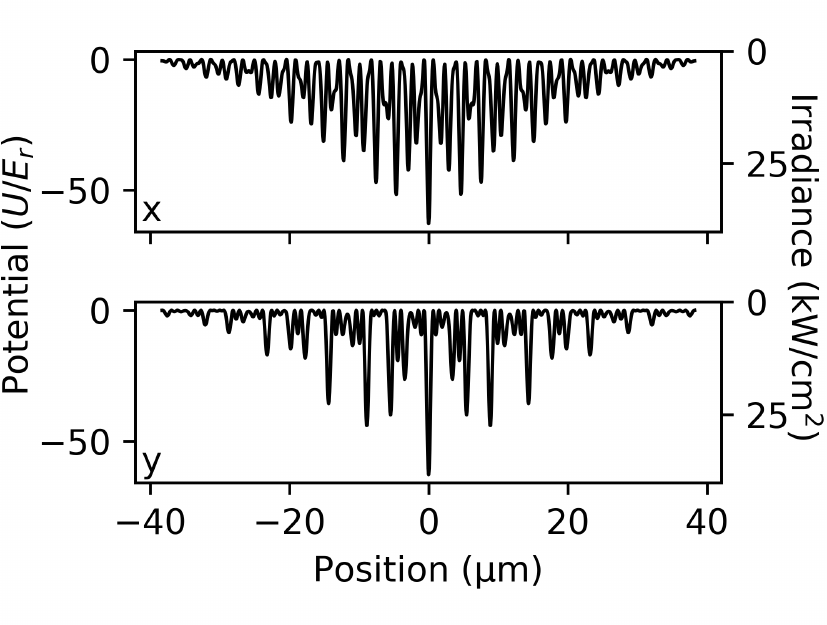}
	\caption{Line cuts along the $x$-axis (top) and $y$-axis (bottom) of the optical potential in Fig.~\ref{fig:QCsim}, showing the energy scale (left axis), irradiance (right axis, inverted), and position (horizontal axis).  The energy is expressed relative to the atoms' lattice photon recoil energy:
		$E_r/k_B \approx \SI{350}{nK}$, where $k_B$ is Boltzmann's constant.
	}
	\label{fig:cuts}
\end{figure}

\begin{figure}
	\centering%
		\includegraphics{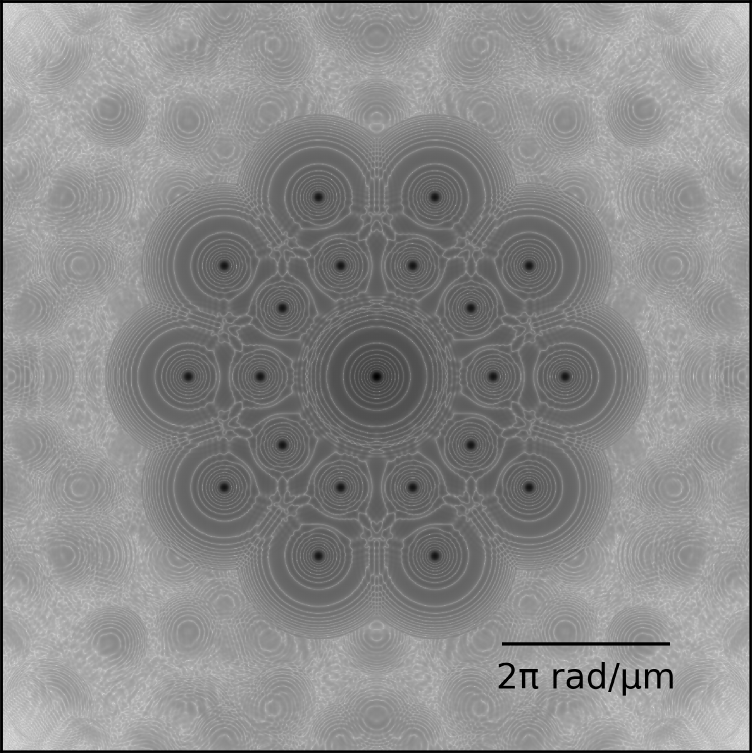}
\caption{Discrete Fourier transform of the optical potential in Fig.~\ref{fig:QCsim}.  The color map is a logarithmic scale with black representing the largest values.  Note the 20 peaks coming from the pairwise differences of the $\mathbf{k}$ vectors and the DC peak, as expected from Eq.~(\ref{eq:Isum}).}\label{fig:fft}
\end{figure}

The irradiance predicted by the simulation is shown in Figs.~\ref{fig:QCsim} and~\ref{fig:cuts}.
For the simulations the laser wavelength is $\SI{850}{nm}$ and the total input power is $\SI{10}{mW}$.
Note that the irradiance is enhanced by a factor of 25 more than a single beam because we are constructively interfering five beams.
The rotational symmetry of the quasicrystal potential is evident in the two-dimensional Fourier transform of the irradiance (Fig.~\ref{fig:fft}).

In normal imaging applications of infinite-conjugate microscopes the distance between the tube lens and objective is unimportant, except that extending this too far will limit the field of view.
In our application, however, the microscope needs to be bi-telecentric, such that the distance between the principle planes of the lenses is equal to the sum of their focal lengths, to ensure that the phase relationships, and therefore the interference pattern, of the light projected onto the atoms' plane matches those where the beams from the fibers initially intersect.
Otherwise, additional distortion and field curvature aberrations will deform the potential, creating unwanted effective strain in the QC lattice\cite{Tian2015,Fujita2016}.

For maximal interference contrast, all of the beams must have the same polarization.
For two interfering beams, such as Eq.~(\ref{eq:pancakeI}), the solution is for the beams to be linearly polarized perpendicular to their common plane of propagation.
For the QC projection system, there is no common plane, so it is not possible to match the polarizations of the 5 beams.
The optimal choice is for each beam to be circularly polarized.
In this configuration a tiny fraction of the optical power ($\sim\sin^2\alpha$, where $\alpha=\SI{10}{mrad}$ is the angle each beam makes with the optical axis) will form a traveling wave with longitudinal polarization relative to the optical axis, but the remainder combines coherently to form the standing wave interference pattern in the transverse plane.
This traveling wave component contributes negligible heating to the atoms.

The potential energy landscape of atoms in the optical field is approximately
\begin{equation}\label{eq:acstark}
U(\mathbf{r}) = \frac{3\pi c^2}{2 \omega_0^3}\,\frac{\Gamma}{\Delta}\,I(\mathbf{r}),
\end{equation}
where $I(\mathbf{r})$ is the position-dependent irradiance of the laser,  $\Delta$ is the difference in frequency (detuning) between the laser and the nearest optical transition, $\omega_0$\cite{Miller1993}.
The equation above assumes $|\Delta| >> \Gamma$, so that hyperfine effects and power broadening may be ignored\cite{Foote2005}.
Note that while the irradiance is necessarily non-negative, the detuning may have either sign, resulting in either an attractive or repulsive potential for red- and blue-detuned light, respectively.
In our experiment we exclusively use red-detuned, therefore attractive, optical trapping potentials.
For our simulation, we are assuming \textsuperscript{87}Rb atoms.
The relevant nearby transition is the D1 line at \SI{795}{nm} ($\omega_0=2\pi\times (\SI{377}{THz})$) with a linewidth of $\Gamma = 2\pi \times (\SI{5.75}{MHz})$\cite{steck-rb87}.
For our simulations the assumed laser wavelength is $\lambda=\SI{850}{nm}$, giving a detuning value of $\Delta = 2\pi\times(\SI{-24.4}{THz}) \approx \num{-4.2e6}\Gamma$.

The maximum potential depth at the atoms, given by Eq.~\eqref{eq:acstark}, is about $U=\SI{-20}{\micro\kelvin}\; k_B$ or $-60$ lattice photon recoil energies, $E_r = {(h/\lambda)}^2/m \approx \SI{350}{nK}\,k_B$.
Because we are constructively interfering five beams, we get an intensity enhancement of $5^2$ in each well.
The tunneling time between adjacent sites is long, on the order of 
\SI{0.5}{s} (approximating adjacent sites as a quartic double-well potential\cite{Verguilla-Berdecia1993}), but this can be decreased by reducing the overall lattice depth.
The detuning of the lattice laser is not large enough to neglect heating from spontaneous emission, which occurs at a rate of \cite{Foote2005}
\[
\Gamma_{\textrm{scat.}} \approx
\frac{3\pi c^2}{2 \hbar\omega_0^3}\,\frac{\Gamma^2}{\Delta^2}\,I(\mathbf{r}).
\]
From this, the heating of the atoms is estimated to be about $\SI{0.3}{\micro\kelvin\per\second}$.

The two-dimensional spatial Fourier transform of the irradiance (Fig.~\ref{fig:fft}) shows the presence of 20 Fourier peaks plus the DC peak, as anticipated by Eq.~(\ref{eq:Isum}).
This is one distinction between our artificial potential and a real quasicrystal.
Real quasicrystals show an infinite sequence of quasi-Bragg peaks of wavevectors with irrational scaling\cite{Janot2012}.
The peaks missing from our system are at large wavevectors and therefore describe short-range, high-energy modes.
We are more interested in long-range, low-energy excitations; our system will be adequate for studying low-temperature behavior.

\begin{figure}
	\centering
	\includegraphics{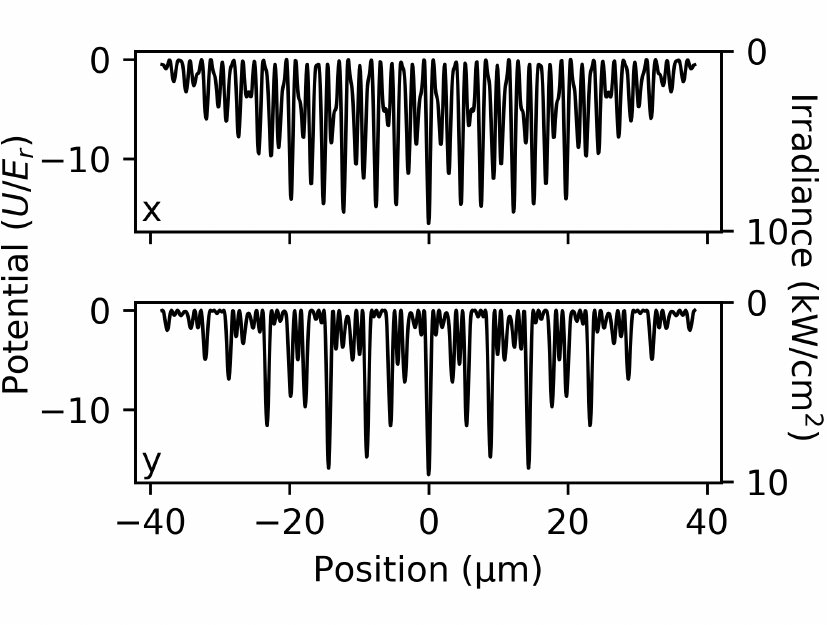}
	\caption{Central cuts of the apodized quasicrystal lattice simulation, showing a flattened potential envelope at the cost of overall trap depth.  Compare with Fig.~\ref{fig:cuts}.}\label{fig:apod}
\end{figure}

The central part of the potential is deeper because of the gaussian profile of the individual beams.
We can flatten out the profile by inserting a neutral density apodizing filter (e.g.~inverse-gaussian filters commercially available from Reynard\cite{reynard} or approximately generated by imaging a gaussian beam onto a negative photographic plate) in the beam path at the ``object'' plane of the microscope (see Fig.~\ref{fig:layout}), at a cost of overall trap depth. 
We have simulated this configuration with an inverse-gaussian filter with maximum attenuation of 0.3 OD and $1/e^2$ radius of $\SI{1.2}{mm}$, showing a flat potential envelope with a diameter of approximately \SI{40}{\micro\meter} (Fig.~\ref{fig:apod}), compared to Fig.~\ref{fig:QCsim}.
The remaining site-to-site variations in depth after the flattening the potential envelope are inherent to quasicrystals.
Our particular geometry has eight types of lattice sites, depending on the configuration of each site's nearest neighbors\cite{Namin2016}.
In particular, the number of nearest neighbors ranges from three to seven, resulting in varying well depths and tunneling rates.

The relative phases between the 5 input beams must be stable over a bandwidth comparable to the classical oscillation frequency within the individual lattice cites to avoid parametric heating\cite{Savard1997}, but the absolute values are not critical. 
Applying relative phases will alter the quasicrystal pattern (e.g.~the individual frames of Fig.~\ref{fig:strip}), but the quasicrystal symmetry itself is not dependent on having specific values of the phases.
If the phases are intentionally shifted smoothly, the lattice potential evolves by exchanging nonlocal pairs of high and low points.
This excitation is called a phason and is unique to quasicrystals. 
Further discussion of phasons follows in Sec.~\ref{sec:mod}.

The optical configuration is relatively robust to alignment errors of the fiber locations.
Although it is difficult to quantify how these errors affect the potential, a reasonable qualitative evaluation is to look at the resulting Fourier transforms of the potential.
We simulated randomly displacing the fibers in the transverse directions and found that for mean displacement errors $<\SI{200}{\micro\meter}$ we still get 21 distinct Fourier peaks in two rings of 10 plus the DC peak.
For larger displacement errors, the peaks begin to split, the ring structures in Fourier space break, and the resulting potentials develop striped structures.
These tolerance simulation results make some intuitive sense: the displacement error of the beams must be less than the beams' $1/e^2$ radius, which is also $\SI{200}{\micro\meter}$.
We note that the potential may be monitored by directly imaging the beams, so any alignment errors can be readily corrected.

\section{Proof-of-concept demonstration}\label{sec:mask}

\begin{figure*}
	\centering
	\includegraphics[width=0.9\textwidth]{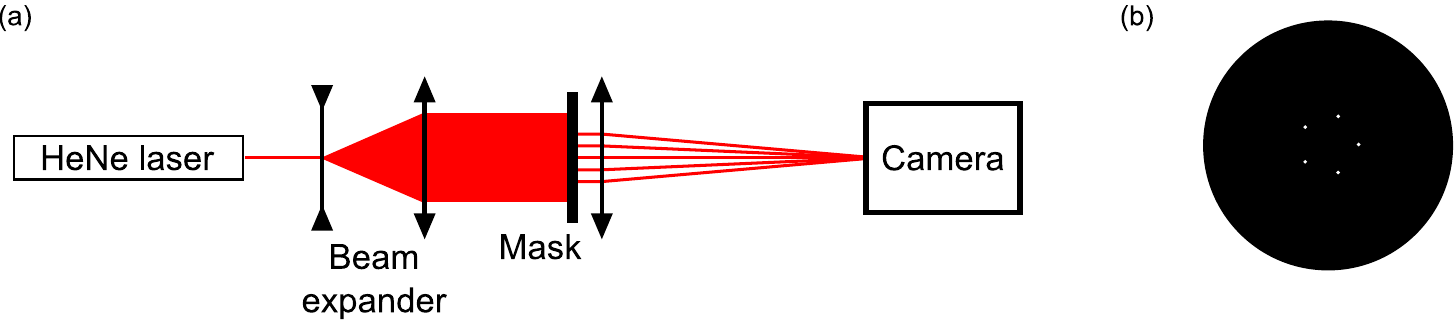}
	\caption{Simplified system for demonstrating quasicrystal optical potential
	(a) Schematic (not to scale) of the system.
	The source laser is a 2-mW HeNe laser operating with a single longitudinal mode and $\mathrm{TEM}_{00}$ profile with $1/e^2$ radius of \SI{1}{mm}.
	The beam is enlarged with a $5\times$ beam expander before striking the mask.
	The focusing lens after the mask has a nominal focal length of \SI{500}{mm}, and the camera (monochromatic CCTV camera with lens removed) is positioned at the back focal plane of this lens.
	The camera also has a slight tilt (not shown) to mitigate etalon fringes caused by the camera sensor window.
	(b) Mask (reduced scale) used to generate the quasicrystal image.
	The 5 holes are equally spaced around a circle of radius \SI{3}{mm} and each has an approximate diameter of \SI{0.2}{mm}.
	The mask is mounted in a fixed \SI{25.4}{mm} lens mount.
	}%
	\label{fig:mockup}
\end{figure*}

As a proof-of-concept demonstration of our method, we assembled a low-power system from a 2-mW HeNe laser, shadow mask approximating the apertures of the five collimation lenses, and a 500-mm focusing lens (Fig.~\ref{fig:mockup}).
The shadow mask was made by poking holes into a piece of paper with a pin.

\begin{figure}
		\centering%
		\includegraphics[height=2.0in]{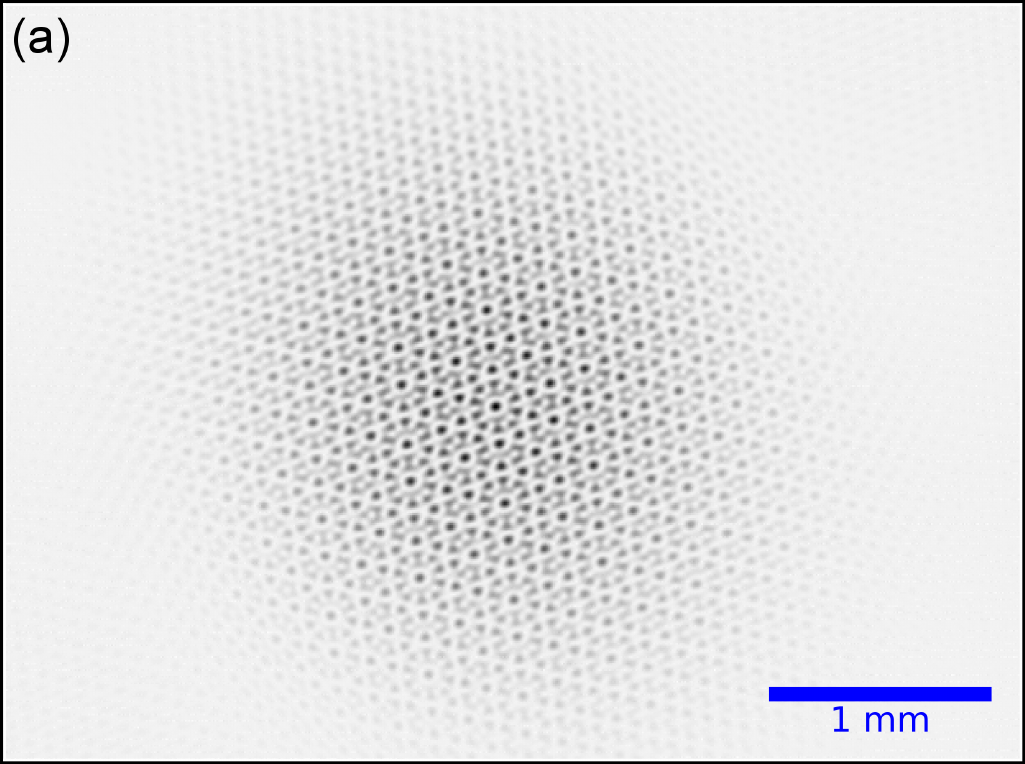}
		\includegraphics[height=2.0in]{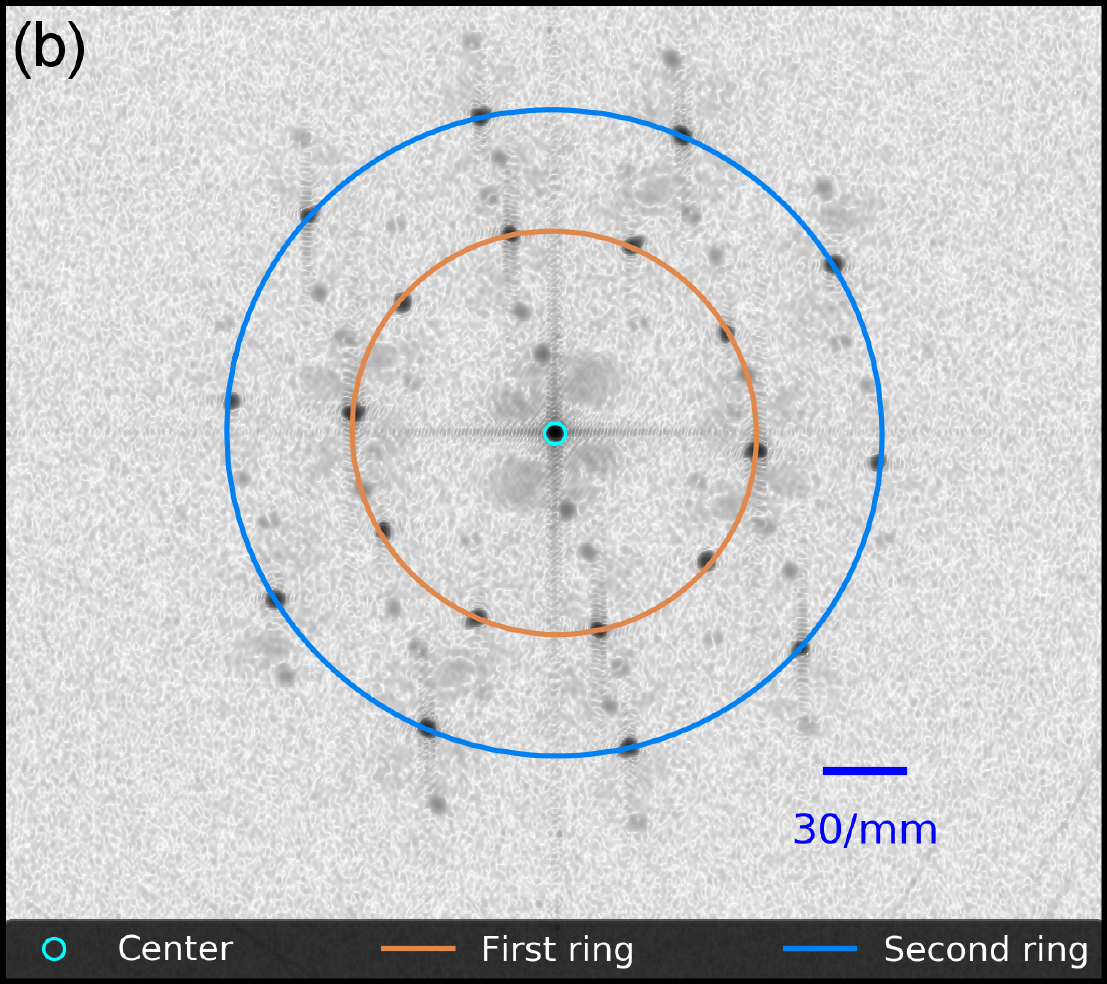}%
		\caption{
			(a) Image and (b) Fourier transform of the image generated by the proof-of-concept system described in Fig.~\ref{fig:mockup}.
			The Fourier image is overlaid with ellipses fit to the locations of the two rings of peaks, showing that the locations agree with the predicted values (see Table~\ref{tab:FFTfit}).
			The fit includes the intentional tilt of the imaging camera. 
			Compare the images here to the simulations in Figs.~\ref{fig:QCsim} and~\ref{fig:fft}.
		}\label{fig:HeNe}
\end{figure}

The interference pattern at the back focal point of the lens was imaged using a CCD camera (Fig.~\ref{fig:HeNe}a).
To verify the 5-fold rotational symmetry of the optical potential, we calculated the two-dimensional discrete Fourier transform of the image (Fig.~\ref{fig:HeNe}b).
The peaks are arranged in two rings of 10 peaks, consistent with 5-fold rotational symmetry, and
the spatial frequencies of the peaks (Table~\ref{tab:FFTfit}) are consistent with the difference of the $\mathbf{k}$-vectors of the incident rays, as predicted by Eq.~\eqref{eq:Isum}.
The ratio of the spatial frequencies of the outer ring of peaks and the inner ring is equal to the golden mean $(1+\sqrt{5})/2 \approx 1.618$, as predicted for regular quasicrystals with 5-fold rotational symmetry\cite{Janot2012,Janssen2018}.
Note that this mask method of generating the optical potentials is not suitable for use in the actual experiment because it is extremely inefficient with the laser power; the power throughput is only 0.1\%.

\begin{table}
	\begin{center}
\caption{Fourier transform fit results}\label{tab:FFTfit}
\begin{tabular}{p{3cm}cc}
Quantity & Predicted & Fitted Values \\ \hline
In-plane wavevector of inner ring ($k_\text{in}$) & $\SI{70.0}{rad/mm}$ & $\SI{74.1}{rad/mm}$ \\
In-plane wavevector of outer ring ($k_\text{out}$)& $\SI{113.2}{rad/mm}$ & $\SI{119.3}{rad/mm}$ \\
Ratio of $k$-vectors ($k_\text{out}/k_\text{in}$) & $(1+\sqrt{5})/2 \approx 1.618$ & $1.610$ \\
Camera rotation about vertical axis & --- & $10.1^\circ $
\end{tabular}
\end{center}
\end{table}

\section{Modulation of quasicrystal potentials}\label{sec:mod}
\begin{figure}
   \centering
   \includegraphics[scale=0.35]{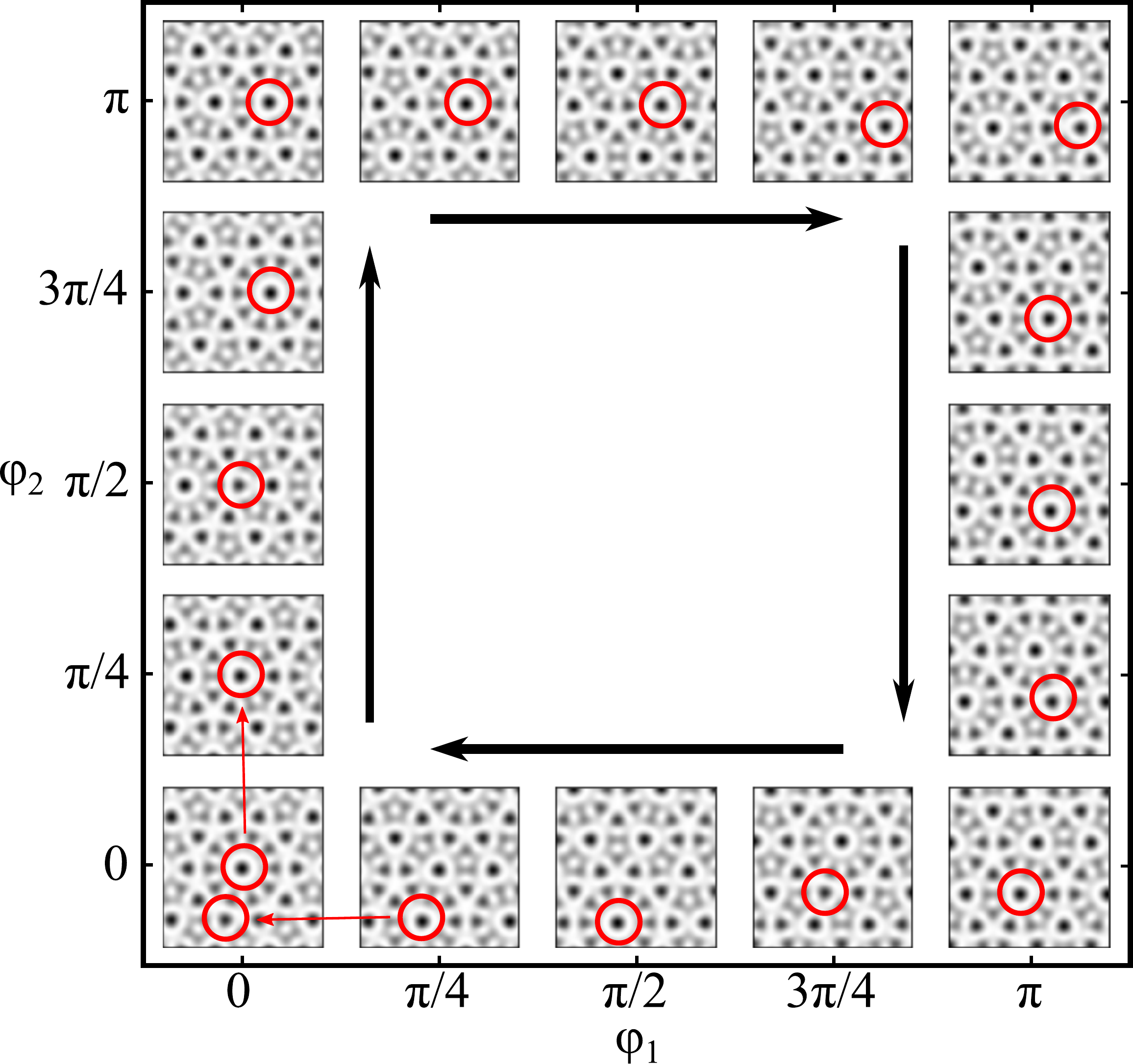}
   \caption{
	   Simulation of a phason modulation and transport.  The relative phases of two of the laser beams ($\phi_{1,2}$), corresponding to the horizontal and vertical location of each frame, are slowly changed in a cycle as time evolves clockwise in the diagram, starting from the bottom left frame.
	   The phases return to their initial values after the cycle, but the locations of the atoms will have shifted.
	   The red circles highlight the evolution of one potential well, which
	   evolves to a new position, an example of geometric pumping.
	   }%
	   \label{fig:strip}
\end{figure}

Phasons are a type of excitation unique to quasicrystals and to date unstudied in cold atoms systems.
They can be created in our system by modulating the relative phases of the generating laser beams.  As shown in Fig.~\ref{fig:strip}, changing the phase of the beams generates a discontinuous translation of the potential pattern.
However, unlike in periodic lattices, such as a 1D lattice, Eq.~(\ref{eq:pancakeI}), the translation does not happen in physical space but in the higher-dimensional configuration space\cite{Socolar1986}.
Phason excitations mathematically fulfill a similar role to that phonons typically play in periodic systems.
Unlike phonons, phasons represent nonlocal correlations between lattice sites as the lattice sites in the fictitious configuration space slide in and out of the cut representing physical space.
In material systems, phasons are responsible for the diffuse background in x-ray scattering experiments\cite{Letoublon2001,Boissieu2005}.

Controlling the relative phases of the interfering beams in real-time, for example using electro-optic modulators on each fiber (Fig.~\ref{fig:layout}) or by modulating the temperatures or strains of the fibers\cite{Lagakos1981}, will generate phason excitations while the atoms are present.
These modes are not directly accessible in solid materials, so measuring of them will provide a novel probe into quasicrystal physics.
Modulating two or more phases in a closed cycle can result in net quantized translation of the particles, known as geometric or Thouless pumping\cite{Thouless1983}.
Thouless pumping has recently been observed in other cold atom systems with topological order\cite{Ma2017,Bandres2016,Levy2015} and in particular bichromatic lattices, a type of one-dimensional quasicrystal\cite{Lohse2015,Nakajima2016,Lu2016}.
This behavior is indicative of the underlying topological order of the system and can be used to empirically determine the Chern numbers of the quasicrystal\cite{Kraus2012}.
Moving beyond adiabatic changes to the potential, rapid cyclic modulation of the phason modes may also induce topologically protected transport modes in the quasicrystal\cite{Bandres2016}.

Figure~\ref{fig:strip} shows a simulation of an adiabatic phason excitation.
The phases of two of the beams are shifted in a cycle such that the final phase is the same as the initial phase, notably \emph{without} invoking any $2\pi$ periodicity.
Note that the circled well divides several times through the sequence, 
and that a particle that started in the initially highlighted well will have its wavefunction split and merged during these divisions, resulting in quantum entanglement of the wells in a way analogous to photons traversing a series of beamsplitters.

\section{Conclusion}
By utilizing nearly co-propagating beams, we have demonstrated the creation of 2D nonperiodic optical potentials that model quasicrystal behavior.  Our system is compact and gives the user control of the lattice scale through the tilt angle of the beams.  We also described how to create phason excitations though controlled phase modulation of the constituent beams.  Furthermore, cycling of the beam phases will allow geometric pumping and measurement of topological parameters through transport measurements.

\section*{Funding}
The Charles E. Kaufman Foundation, a member organization of the Pittsburgh Foundation (KA2015-79202). 





\bibliography{QX-AO}

\end{document}